\documentclass[aps,prl,twocolumn,showpacs,superscriptaddress]{revtex4-1}

\usepackage{amsfonts,mathrsfs,amsmath,amsthm,amssymb}
\usepackage{bm,times}
\usepackage{graphicx,epsfig}
\usepackage[colorlinks=true,breaklinks=true,linkcolor=blue,citecolor=blue,urlcolor=blue]{hyperref}

\def \non {\nonumber}

\newcommand{\be}{\begin{eqnarray}}
\newcommand{\ee}{\end{eqnarray}}

\makeatletter
    
    \newcommand{\Rmnum}[1]{\expandafter\@slowromancap\romannumeral #1@}
\makeatother

\begin{document}
\title{New construction of nine-qubit error-correcting code}
\author{Long Huang}
\email{huangl@sicnu.edu.cn}
\affiliation{Coll Phys \& Elect Engn, Sichuan Normal University, Chengdu 610101, China}
\author{Xiaohua Wu}
\email{wxhscu@scu.edu.cn}
\affiliation{College of Physical Science and Technology, Sichuan University, Chengdu 610064, China}

\begin{abstract}
We report new construction of nine-qubit error-correcting code, which introduces two new nine-qubit codes and one new three-qubit code. Because both the new two nine-qubit codes have the normal logical operators, as opposed to the nine-qubit Shor code, it results in different performance when the three codes are applied in concatenated quantum error-correction. On the other hand, one of the two nine-qubit codes has the same stabilizer generators as the nine-qubit Shor code, they are more suitable for the high-wight bit-flip noise, and the other code has the different stabilizer generators, which is more suitable for the high-wight phase-flip noise. This work is enlightening to the construction of quantum error-correcting codes, and adds more options for optimizing the performance of quantum error-correction.
\end{abstract}

\date{\today}

\maketitle

In quantum computation and communication, quantum error-correction (QEC) developed from classic schemes to preserving coherent states from noise and other unexpected interactions. It was independently discovered by Shor and Steane~\cite{Shor,Steane}. The QEC conditions were proved independently by Bennett, DiVincenzo, Smolin and Wootters~\cite{Bennett}, and by Knill and Laflamme~\cite{Knill}. QEC codes are introduced as active error-correction. The nine-qubit code was discovered by Shor, called the Shor code. The seven-qubit code was discovered by Steane, called the Steane code. The five-qubit code was discovered by Bennett, DiVincenzo, Smolin and Wootters~\cite{Bennett}, and independently by Laflamme, Miquel, Paz and Zurek~\cite{Laflamme}.

There are many constructions for specific classes of quantum codes. Rains, Hardin, Shor and Sloane~\cite{Rains} have constructed interesting examples of quantum codes lying outside the stabilizer codes. Gottesman~\cite{Gottesman2} and Rains~\cite{Rains2} construct non-binary codes and consider fault-tolerant computation with such codes. Aharonov and Ben-Or~\cite{Aharonov} construct non-binary codes using interesting techniques based on polynomials over finite fields, and also investigate fault-tolerant computation with such codes. Approximate QEC can lead to improved codes was shown by Leung, Nielsen, Chuang and Yamamoto~\cite{Leung}.

Calderbank and Shor~\cite{Calderbank}, and Steane~\cite{Steane2} used ideas from classical error-correction to develop the CSS (Calderbank-Shor-Steane) codes. Calderbank and Shor also stated and proved the Gilbert-Varshamov bound for CSS codes. Gottesman~\cite{Gottesman} invented the stabilizer formalism, and used it to define stabilizer codes, and investigated some of the properties of some specific codes. Independently, Calderbank, Rains, Shor and Sloane~\cite{Calderbank2} invented an essentially equivalent approach to QEC based on ideas from classical coding theory.

QEC codes are introduced as active error-correction. Another way, passive error-avoiding techniques contain decoherence-free subspaces~\cite{Duan,Lidar,Zanardi} and noiseless subsystem~\cite{KandLV,Zanardi2,Kempe}. Recently, it has been proven that both the active and passive QEC methods can be unified~\cite{Kribs,Poulin 05,Kribs2}. So, more QEC codes means more options for suppressing noise, and more options for optimizing the performance of QEC. Meanwhile, the quantum resources required are also different for codes with different constructions, as the two five-qubit codes in Ref.~\cite{Bennett,Laflamme}. Inspired by the two five-qubit codes, and motivated by the question that whether other error-correcting codes also have the corresponding codes. Inspired by the phenomenon that the logical operators of the nine-qubit Shor code are the reverse of what one might expect, and we are also motivated by the question that is there one nine-qubit code which has the normal logical operators.

In this work, we finally got the answer to the original questions. We constructed two new nine-qubit codes, and both the two nine-qubit codes have the normal logical operators, as opposed to the nine-qubit Shor code. As shown in Fig.~\ref{figure2}, the two nine-qubit codes are constructed with opposite order of outer and inner encoding. One has the same stabilizer generators as the nine-qubit Shor code, and the other has the different stabilizer generators. As shown in Table~(\ref{t1}) and~(\ref{t2}), we can obtain concatenated QEC protocols which have better performance than the concatenated QEC protocols with the Shor code.

\begin{figure}[tbph]
\centering
\includegraphics[width=0.4 \textwidth]{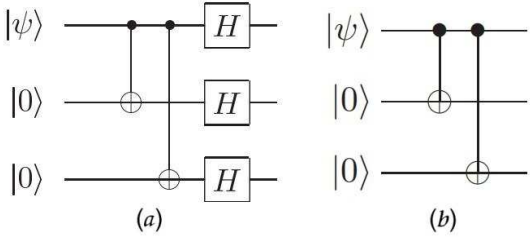}\\
\caption{(a) Encoding circuits for the phase-flip code, and it is applied as the outer encoding in the nine-qubit Shor code.
 (b) Encoding circuits for the bit-flip code, and it is applied as the inner encoding in the nine-qubit Shor code. }\label{figure1}
\end{figure}

\begin{figure}[tbph]
\centering
\includegraphics[width=0.4 \textwidth]{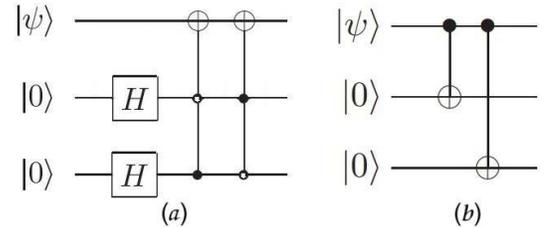}\\
\caption{(a) Encoding circuits for another phase-flip code, it is applied as the outer encoding for the new nine-qubit code in Eq.~(\ref{e2}), and as the inner encoding for the new nine-qubit code in Eq.~(\ref{e4}).
 (b) Encoding circuits for the bit-flip code, it is applied as the inner encoding for the new nine-qubit code in Eq.~(\ref{e2}), and as the outer encoding for the new nine-qubit code in Eq.~(\ref{e4}). }\label{figure2}
\end{figure}

We begin with a short review of the nine-qubit Shor code as presented in~\cite{Shor,Nielsen}. The code is a combination of the three-qubit phase-flip and bit-flip codes. As depicted in Fig.~\ref{figure1}~(a), the sate $|\psi\rangle=a|0\rangle+b|1\rangle$ is encoded to $|\tilde{\psi}\rangle=a|0_\mathcal{L}\rangle+b|1_\mathcal{L}\rangle$,
\be
\label{e1}
|0_\mathcal{L}\rangle=\frac{1}{\sqrt{8}}[|000\rangle+|001\rangle+|010\rangle+|011\rangle\non\\
+|100\rangle+|101\rangle+|110\rangle+|111\rangle],\non\\
|1_\mathcal{L}\rangle=\frac{1}{\sqrt{8}}[|000\rangle-|001\rangle-|010\rangle+|011\rangle\non\\
-|100\rangle+|101\rangle+|110\rangle-|111\rangle].
\ee
The code in Eq.~(\ref{e1}) is known as the three-qubit phase-flip code, and the errors \{$I$, $Z_1$, $Z_2$, $Z_3$\} with the stabilizer $\langle X_1X_2, X_2X_3\rangle$. The logical $X$ operator is $Z_1Z_2Z_3$, and the logical $Z$ operator is one of $X_1X_2X_3$, $X_1$, $X_2$, and $X_3$. Then, the nine-qubit Shor code can be constructed by encoding every $|0\rangle$ and $|1\rangle$ in Eq.~(\ref{e1}) to $|000\rangle$ and $|111\rangle$. Accordingly, the logical $X$ and $Z$ operators are also expanded on nine-qubit, which are $Z_1Z_2Z_3Z_4Z_5Z_6Z_7Z_8Z_9$/$Z_{O_1}Z_{O_2}Z_{O_3}$ ($O_1$ is one of qubits 1-3, $O_2$ is one of qubits 4-6, $O_3$ is one of qubits 7-9) and one of $X_1X_2X_3X_4X_5X_6X_7X_8X_9$, $X_1X_2X_3$, $X_4X_5X_6$, and $X_7X_8X_9$.

On the other hand, we found that the code words of other codes can be obtained by applying all combinations from subgroup of stabilizer generators on the states $|0\rangle^{\otimes n}$ and $|1\rangle^{\otimes n}$, and $n$ is the number of qubits, examples are shown in Appendix~A. We can obtain one new nine-qubit code through this way, and the code words are,
\be
\label{e2}
|0_\mathcal{L}\rangle
&=&\frac{1}{2}[|000000000\rangle+|111111000\rangle\non\\
&&+|000111111\rangle+|111000111\rangle],\non\\
|1_\mathcal{L}\rangle
&=&\frac{1}{2}[|111111111\rangle+|000000111\rangle\non\\
&&+|111000000\rangle+|000111000\rangle].
\ee
This nine-qubit code can protect against arbitrary single-qubit errors as the nine-qubit Shor code. On the other hand, the difference is the logical Pauli $Z$ operator of this code is $Z_1Z_2Z_3Z_4Z_5Z_6Z_7Z_8Z_9$/$Z_{O_1}Z_{O_2}Z_{O_3}$, and the logical Pauli $X$ operator is one of $X_1X_2X_3X_4X_5X_6X_7X_8X_9$, $X_1X_2X_3$, $X_4X_5X_6$, and $X_7X_8X_9$.

Similar to the nine-qubit Shor code, this nine-qubit code also can be constructed as shown in Fig.~\ref{figure2}. The outer three-qubit code is one new code, which is depicted in Fig.~\ref{figure2}~(a). With this code, the sate $|\psi\rangle=a|0\rangle+b|1\rangle$ is encoded to $|\tilde{\psi}\rangle=a|0_\mathcal{L}\rangle+b|1_\mathcal{L}\rangle$,
\be
\label{e3}
|0_\mathcal{L}\rangle=\frac{1}{2}[|000\rangle+|011\rangle+|101\rangle+|110\rangle],\non\\
|1_\mathcal{L}\rangle=\frac{1}{2}[|111\rangle+|100\rangle+|010\rangle+|001\rangle].
\ee
The three-qubit code in Eq.~(\ref{e3}) is also one three-qubit phase-flip code, and the errors \{$I$, $Z_1$, $Z_2$, $Z_3$\} with the stabilizer $\langle X_1X_2, X_2X_3\rangle$. The logical $Z$ operator is $Z_1Z_2Z_3$, and the logical $X$ operator is one of $X_1X_2X_3$, $X_1$, $X_2$, and $X_3$. On the other hand, this three-qubit code also can be obtained from the subgroup \{$I$, $X_1X_2$, $X_2X_3$\},
\be
|0_\mathcal{L}\rangle&=&\frac{1}{2}(III+X_1X_2+X_2X_3+X_1X_3)|000\rangle\non\\
&=&\frac{1}{2}[|000\rangle+|011\rangle+|101\rangle+|110\rangle],\non\\
|1_\mathcal{L}\rangle&=&\frac{1}{2}(III+X_1X_2+X_2X_3+X_1X_3)|111\rangle\non\\
&=&\frac{1}{2}[|111\rangle+|100\rangle+|010\rangle+|001\rangle].\non
\ee
Through comparison, the construction of this three-qubit code is more similar to the three-qubit bit-flip code, whose logical $X$ operator is $X_1X_2X_3$, and the logical $Z$ operator is one of $Z_1Z_2Z_3$, $Z_1$, $Z_2$, and $Z_3$.

Then, we will introduce quantum error-correcting of this nine-qubit code. Because this nine-qubit code can be constructed with the concatenation of bit-flip code and the three-qubit code in Eq.~(\ref{e3}), we begin with the error-correcting of the two three-qubit codes.

For the bit-flip code, the sate $|\psi\rangle=a|0\rangle+b|1\rangle$ is encoded to $|\tilde{\psi}\rangle=a|000\rangle+b|111\rangle$. When bit-flip error occurred on one qubit, measuring with the stabilizer $\langle Z_1Z_2, Z_2Z_3\rangle$,
\be
X_1|\tilde{\psi}\rangle&=&a|100\rangle+b|011\rangle,S_1=-1,1;\non\\
X_2|\tilde{\psi}\rangle&=&a|010\rangle+b|101\rangle,S_2=-1,-1;\non\\
X_3|\tilde{\psi}\rangle&=&a|001\rangle+b|110\rangle,S_3=1,-1.\non
\ee
Here, $S_n$ is the measurement result for the n-qubit. We should notice that if bit-phase-flip error occurred on one qubit, the measurement results with the stabilizer $\langle Z_1Z_2, Z_2Z_3\rangle$ are
\be
Y_1|\tilde{\psi}\rangle&=&i[a|100\rangle-b|011\rangle],S_1=-1,1;\non\\
Y_2|\tilde{\psi}\rangle&=&i[a|010\rangle-b|101\rangle],S_2=-1,-1;\non\\
Y_3|\tilde{\psi}\rangle&=&i[a|001\rangle-b|110\rangle],S_3=1,-1.\non
\ee
The measurement results for each qubit occurred one-qubit bit-flip or bit-phase-flip errors are the same, which means the bit-flip code can be used to correct one-qubit bit-flip or bit-phase-flip error in single-error environment, and not in mixed noise environment because the bit-flip and bit-phase-flip errors can not be identified simultaneously.

For the three-qubit code in Eq.~(\ref{e3}), the sate $|\psi\rangle=a|0\rangle+b|1\rangle$ is encoded to $|\tilde{\psi}\rangle=a[|000\rangle+|011\rangle+|101\rangle+|110\rangle]+b[|111\rangle+|100\rangle+|010\rangle+|001\rangle]$. When phase-flip error occurred on one qubit, measuring with the stabilizer $\langle X_1X_2, X_2X_3\rangle$,
\be
Z_1|\tilde{\psi}\rangle&=&a[|000\rangle+|011\rangle-|101\rangle-|110\rangle]\non\\
&&+b[-|111\rangle-|100\rangle+|010\rangle+|001\rangle],\non\\
&&R_1=-1,1;\non\\
Z_2|\tilde{\psi}\rangle&=&a[|000\rangle-|011\rangle+|101\rangle-|110\rangle]\non\\
&&+b[-|111\rangle+|100\rangle-|010\rangle+|001\rangle],\non\\
&&R_2=-1,-1;\non\\
Z_3|\tilde{\psi}\rangle&=&a[|000\rangle-|011\rangle-|101\rangle+|110\rangle]\non\\
&&+b[-|111\rangle+|100\rangle+|010\rangle-|001\rangle],\non\\
&&R_3=1,-1.\non
\ee
Here, $R_n$ is the measurement result for the n-qubit. And if bit-phase-flip error occurred on one qubit, the measurement results with the stabilizer $\langle X_1X_2, X_2X_3\rangle$ are
\be
Y_1|\tilde{\psi}\rangle&=&i[a[|100\rangle+|111\rangle-|001\rangle-|010\rangle]\non\\
&&+b[-|011\rangle-|000\rangle+|110\rangle+|101\rangle]],\non\\
&&R_1=-1,1;\non\\
Y_2|\tilde{\psi}\rangle&=&i[a[|010\rangle-|001\rangle+|111\rangle-|100\rangle]\non\\
&&+b[-|101\rangle+|110\rangle-|000\rangle+|011\rangle]],\non\\
&&R_2=-1,-1;\non\\
Y_3|\tilde{\psi}\rangle&=&i[a[|001\rangle-|010\rangle-|100\rangle+|111\rangle]\non\\
&&+b[-|110\rangle+|101\rangle+|011\rangle-|000\rangle]],\non\\
&&R_3=1,-1.\non
\ee
The measurement results for each qubit occurred one-qubit phase-flip or bit-phase-flip errors are the same, which means the three-qubit code can be used to correct one-qubit phase-flip or bit-phase-flip error in single-error environment, and not in mixed noise environment because the phase-flip and bit-phase-flip errors can not be identified simultaneously.

Fortunately, the nine-qubit code Eq.~(\ref{e2}) can be constructed with the concatenation of the two three-qubit codes, which making the error correction of single-qubit error possible. As shown in the code words Eq.~(\ref{e2}), the inner three qubits 1-3, 4-6, or 7-9 are encoded with bit-flip code, and the outer three qubits ($O_1$ is one of qubits 1-3, $O_2$ is one of qubits 4-6, $O_3$ is one of qubits 7-9) are encoded as Eq.~(\ref{e3}). We can test and recover any one-qubit error through the measurement with the two series stabilizer.

For example, to test if any error occurred on the first qubit. We begin the measurement with the stabilizer $\langle Z_1Z_2, Z_2Z_3\rangle$ on qubits 1-3. (I) If $S_1=1,1$, which means no bit-flip or bit-phase-flip error occurred on the first qubit, then we continue the measurement with the stabilizer $\langle X_1X_2X_3X_4X_5X_6, X_4X_5X_6X_7X_8X_9\rangle$. The measurement result $R_1=1,1$ means no error occurred on the first qubit, and $R_1=-1,1$ means phase-flip error occurred on the first qubit. (II) If $S_1=-1,1$, which means bit-flip or bit-phase-flip error occurred on the first qubit, then we continue the measurement with the stabilizer $\langle X_1X_2X_3X_4X_5X_6, X_4X_5X_6X_7X_8X_9\rangle$. The measurement result $R_1=1,1$ means bit-flip error occurred on the first qubit, and $R_1=-1,1$ means bit-phase-flip error occurred on the first qubit (there are three possible cases, which are bit-flip error occurred on qubit-1, and phase-flip error occurred on qubit-1/qubit-2/qubit-3, but to apply bit-phase-flip operation on qubit-1 is enough for recovering the three cases).

In fact, we found most errors occurred on qubits $O_1O_2O_3$ can be recovered, except few two-qubit and three-qubit errors. The proof is as follows, we begin the measurements from the inner qubits. To measure qubits 1-3 with the stabilizer $\langle Z_1Z_2, Z_2Z_3\rangle$, qubits 4-6 with the stabilizer $\langle Z_4Z_5, Z_5Z_6\rangle$, and qubits 7-9 with the stabilizer $\langle Z_7Z_8, Z_8Z_9\rangle$. Then, we continue the outer measurement with the stabilizer $\langle X_1X_2X_3X_4X_5X_6, X_4X_5X_6X_7X_8X_9\rangle$. Here, we should notice that although the measurements can not distinguish the phase-flip error that occurred on qubit-1/qubit-2/qubit-3, having no influence on recovering them. In the Appendix~B, we give all the measurement results for the errors that occurred on qubits $O_1O_2O_3$, and note which can be recovered.

Based on the three-qubit bit-flip code and the three-qubit code in Eq.~(\ref{e3}), another nine-qubit code can be constructed by exchanging the order of outer and inner encoding, and the code words are,
\be
\label{e4}
|0_\mathcal{L}\rangle=\frac{1}{8}[|000\rangle+|011\rangle+|101\rangle+|110\rangle]^{\bigotimes3},\non\\
|1_\mathcal{L}\rangle=\frac{1}{8}[|111\rangle+|100\rangle+|010\rangle+|001\rangle]^{\bigotimes3}.
\ee
This nine-qubit code can protect against arbitrary single-qubit errors, too. For this code, the logical Pauli $X$ operator of this code is $X_1X_2X_3X_4X_5X_6X_7X_8X_9$/$X_{O_1}X_{O_2}X_{O_3}$, and the logical Pauli $Z$ operator is one of $Z_1Z_2Z_3Z_4Z_5Z_6Z_7Z_8Z_9$, $Z_1Z_2Z_3$, $Z_4Z_5Z_6$, and $Z_7Z_8Z_9$. The stabilizer generators of this code are \{$X_1X_2,X_2X_3,X_4X_5,X_5X_6,X_7X_8,X_8X_9$, $Z_1Z_2Z_3Z_4Z_5Z_6$, $Z_4Z_5Z_6Z_7Z_8Z_9$\}. It results different performance from the code in Eq.~(\ref{e2}) and Shor code, which means it is more suitable for the phase-flip errors. Similarly, we begin the measurements from the inner qubits. To measure qubits 1-3 with the stabilizer $\langle X_1X_2, X_2X_3\rangle$, qubits 4-6 with the stabilizer $\langle X_4X_5, X_5X_6\rangle$, and qubits 7-9 with the stabilizer $\langle X_7X_8, X_8X_9\rangle$. Then, we continue the outer measurement with the stabilizer $\langle Z_1Z_2Z_3Z_4Z_5Z_6, Z_4Z_5Z_6Z_7Z_8Z_9\rangle$. However, the measurements can not distinguish the bit-flip error that occurred on qubit-1/qubit-2/qubit-3, having no influence on recovering them. In the Appendix~C, we give all the measurement results for the errors that occurred on qubits $O_1O_2O_3$, and note which can be recovered.

\begin{table}
\caption{Results of 5-level concatenated QEC, where the initial process matrix $\lambda_{L,L=0}$ is $\lambda_{II}=0.92, \lambda_{XX}=\lambda_{ZZ}=\lambda_{YY}=\frac{0.08}{3}$, and $L$ is the concatenation level of different QEC protocol with the Shor code or the two new codes, respectively.}\label{t1}
\centering\begin{tabular}{p{0.9cm}  p{7.4cm} }
\hline\hline\noalign{\smallskip}
$\lambda-L$ & Shor code $(\lambda_{II},\lambda_{XX},\lambda_{ZZ},\lambda_{YY})$ \\
\hline\noalign{\smallskip}
   $0$  & $(0.92,\frac{0.08}{3},\frac{0.08}{3},\frac{0.08}{3})$    \\
\hline\noalign{\smallskip}
   $1$  & $(0.934261,0.0414538,0.02211,0.00217564)$     \\
\hline\noalign{\smallskip}
   $2$  & $(0.968208,0.0153426,0.0159628,0.00048693)$    \\
\hline\noalign{\smallskip}
  $3$  & $(0.990936,0.00683566,0.00220061,0.0000274494)$     \\
\hline\noalign{\smallskip}
   $4$  & $(0.999438,0.000140051,0.000421678,1.84203\times10^{-7})$     \\
\hline\noalign{\smallskip}
   $5$  & $(0.999995,4.79366\times10^{-6},1.76974\times10^{-7},0)$      \\
\hline\noalign{\smallskip}
$\lambda-L$     & Codes in Eq.~(\ref{e2}) and~(\ref{e4}) concatenated as (24422)  \\
\hline\noalign{\smallskip}
   $0$      & $(0.92,\frac{0.08}{3},\frac{0.08}{3},\frac{0.08}{3})$  \\
\hline\noalign{\smallskip}
  $1$       & $(0.934261,0.02211,0.0414538,0.00217564)$  \\
\hline\noalign{\smallskip}
  $2$       & $(0.968208,0.0153426,0.0159628,0.00048693)$ \\
\hline\noalign{\smallskip}
  $3$       & $(0.991216,0.00637912,0.00237658,0.0000281788)$ \\
\hline\noalign{\smallskip}
  $4$       & $(0.999471,0.00036764,0.000161228,1.72737\times10^{-7})$  \\
\hline\noalign{\smallskip}
  $5$       & $(0.999998,1.21727\times10^{-6},7.04428\times10^{-7},0)$  \\
\hline\hline
\end{tabular}
\end{table}

\begin{table}
\caption{Results of 5-level concatenated QEC, where the initial process matrix $\lambda_{L,L=0}$ is $\lambda_{II}=0.92, \lambda_{XX}=\lambda_{ZZ}=0.04, \lambda_{YY}=0$, and $L$ is the concatenation level of different QEC protocol with the Shor code or the two new codes, respectively.}\label{t2}
\centering\begin{tabular}{p{0.9cm}  p{7.4cm} }
\hline\hline\noalign{\smallskip}
$\lambda-L$ & Shor code $(\lambda_{II},\lambda_{XX},\lambda_{ZZ},\lambda_{YY})$ \\
\hline\noalign{\smallskip}
   $0$  & $(0.92,0.04,0.04,0)$    \\
\hline\noalign{\smallskip}
   $1$  & $(0.94931,0.0368044,0.0133265,0.000558912)$     \\
\hline\noalign{\smallskip}
   $2$  & $(0.982052,0.00579672,0.0120072,0.000144243)$    \\
\hline\noalign{\smallskip}
  $3$  & $(0.995951,0.00373275,0.000314342,1.98865\times10^{-6})$     \\
\hline\noalign{\smallskip}
   $4$  & $(0.999872,3.04353\times10^{-6},0.000125209,0)$     \\
\hline\noalign{\smallskip}
   $5$  & $(1-4.22988\times10^{-7},4.22988\times10^{-7},0,0)$      \\
\hline\noalign{\smallskip}
$\lambda-L$     & Codes in Eq.~(\ref{e2}) and~(\ref{e4}) concatenated as (24442)  \\
\hline\noalign{\smallskip}
   $0$      & $(0.92,0.04,0.04,0)$  \\
\hline\noalign{\smallskip}
  $1$       & $(0.94931,0.0133265,0.0368044,0.000558912)$  \\
\hline\noalign{\smallskip}
  $2$       & $(0.982052,0.00579672,0.0120072,0.000144243)$ \\
\hline\noalign{\smallskip}
  $3$       & $(0.997705,0.000977668,0.00131379,3.1977\times10^{-6})$ \\
\hline\noalign{\smallskip}
  $4$       & $(0.999958,0.0000259578,0.0000155951,0)$  \\
\hline\noalign{\smallskip}
  $5$       & $(1,0,0,0)$  \\
\hline\hline
\end{tabular}
\end{table}

In this work, we constructed two new nine-qubit codes in~Eq.~(\ref{e2}) and~Eq.~(\ref{e4}), and introduce one new three-qubit code~Eq.~(\ref{e3}). Both the new two nine-qubit codes have the normal logical operators, as opposed to the nine-qubit Shor code, it results in different performance when applied in concatenated QEC. The transversal $X$-gates and $Z$-gates for the Shor code are correspondingly inverted~\cite{Shor,Leung,Huang3}, and not for the codes in~Eq.~(\ref{e2}) and~(\ref{e4}). There may be different applications with these codes in concatenated QEC. When applied in one-level QEC, the code in~Eq.~(\ref{e2}) has the same stabilizer generators as the nine-qubit Shor code, they are more suitable for the high-wight bit-flip noise, and the code in~Eq.~(\ref{e4}) has the different stabilizer generators, which is more suitable for the high-wight phase-flip noise. This work is enlightening to the construction of quantum error-correcting codes, and adds more options for optimizing the performance of QEC. For example, as shown in Table~(\ref{t1}) and~(\ref{t2}), concatenation QEC with the codes~Eq.~(\ref{e2}) and~Eq.~(\ref{e4}) has a better performance than concatenation QEC just with the Shor code, and it is tenable in most cases. And in few cases, the two QEC protocols have the same performance, for example, the first two levels as shown in Table~(\ref{t1}) and~(\ref{t2}).

For fully understanding of the nine-qubit codes, we begin from numbering the bit-flip code $|000\rangle,|111\rangle$ as B-(1), another bit-flip code $\frac{1}{\sqrt{2}}[|000\rangle+|111\rangle], \frac{1}{\sqrt{2}}[|000\rangle-|111\rangle]$ as B-(2), the phase-flip code in Eq.~(\ref{e1}) as P-(1), and the phase-flip code in Eq.~(\ref{e3}) as P-(2). The transformation between the four three-qubit codes can be obtained,
\be
Logical{H}.B-(1)\Rightarrow B-(2),H^{\otimes3}.B-(2)\Rightarrow P-(2),\non\\
Logical{H}.P-(2)\Rightarrow P-(1),H^{\otimes3}.P-(1)\Rightarrow B-(1),\non\\
C(g).B-(1)\Rightarrow P-(2),C(g).B-(2)\Rightarrow P-(1).\non
\ee
Here, $C(g)$ are the all combinations of the stabilizer generators of three-qubit phase-flip code, which list as $\{III+X_1X_2+X_2X_3+X_1X_3\}$. Concatenating one of the bit-flip codes and one of the phase-flip codes, we can obtain four different nine-qubit codes, list as B-(1)P-(1), B-(1)P-(2), B-(2)P-(1), and B-(2)P-(2). And exchange the order of concatenation, we also obtain four different nine-qubit codes, list as P-(1)B-(1), P-(1)B-(2), P-(2)B-(1), and P-(2)B-(2). The eight nine-qubit codes are all possible results we can construct by three-qubit codes. Four of them are effective for correcting any one-qubit errors, which are P-(2)B-(1), the code in Eq.~(\ref{e2}); B-(1)P-(2), the code in Eq.~(\ref{e4}); P-(1)B-(1)=$Logical{H}$.P-(2)B-(1), the Shor code; B-(2)P-(2)=$Logical{H}$.B-(1)P-(2), the code similar with the Shor code. We think the codes P-(2)B-(1)~Eq.~(\ref{e2}) and B-(1)P-(2)~ Eq.~(\ref{e4}) are more basic, because they have the simplest code words and can transform to each other by exchanging the order of concatenation of three-qubit codes. It also indicates that the codes B-(1) and P-(2)~Eq.~(\ref{e3}) are the basic and simplest codes in three-qubit codes, and the inner encoding construction of effective nine-qubit codes must be one of them.

Meanwhile, we note the codes P-(1)B-(1) the Shor code and B-(2)P-(2) have similar performance, but can not transform to each other by exchanging the order of concatenation of three-qubit codes. Noting that we can not construct the nine-qubit codes which are effective for correcting any one-qubit errors and more suitable for the high-wight phase-flip noise without the code P-(2), and it might resulted that we can not reach one better performance than the Shor code as shown in Table~(\ref{t1}) and~(\ref{t2}).

The rest four codes are not effective for correcting any one-qubit errors, and they are just effective for correcting bit-flip error or phase-flip error as the three-qubit codes. B-(1)P-(1) is the code with code words,
\be
|0_\mathcal{L}\rangle=\frac{1}{16\sqrt{2}}[|0\rangle+|1\rangle]^{\bigotimes9},\non\\
|1_\mathcal{L}\rangle=\frac{1}{16\sqrt{2}}[|0\rangle-|1\rangle]^{\bigotimes9};\non
\ee
B-(2)P-(1)=$Logical{H}$.B-(1)P-(1); P-(1)B-(2) is the nine-qubit linear code $|000000000\rangle,|111111111\rangle$; P-(2)B-(2)=$Logical{H}$.P-(1)B-(2).

In realization of QEC, some works have been done~\cite{Kelly,Sohn}. We think he performance of the three-qubit codes and nine-qubit codes is interesting, because the known nine-qubit codes are obtained by concatenating two three-qubit codes. Once the QEC with the three-qubit code had been realized, the realization with the nine-qubit code is nearer. Meanwhile, the nine-qubit codes can be layout with 1D structure or 2D structure in different physical systems, which increases the likelihood of realization. Based on the realization of QEC with nine-qubit codes, any one-qubit error can be corrected, and fault-tolerant quantum computation may be realized in practical quantum computer.

\emph{We thank the reviewers for the helpful comments of~\cite{Huang3}.}

\section{appendix a: The code words obtained from stabilizer generators}
\begin{appendix}
\label{a1}

We found that the code words of other codes can be obtained by applying all combinations from subgroup of stabilizer generators on the states $|0\rangle^{\otimes n}$ and $|1\rangle^{\otimes n}$, and $n$ is the number of qubits.

For the three-qubit bit-flip code, the subgroup is \{$I$, $Z_1Z_2$, $Z_2Z_3$\}, and the code words are,
\be
|0_\mathcal{L}\rangle&=&\frac{1}{2}(III+Z_1Z_2+Z_2Z_3+Z_1Z_3)|000\rangle\non\\
&=&|000\rangle,\non\\
|1_\mathcal{L}\rangle&=&\frac{1}{2}(III+Z_1Z_2+Z_2Z_3+Z_1Z_3)|111\rangle\non\\
&=&|111\rangle.\non
\ee

For the five-qubit code in Ref.~\cite{Bennett}, the subgroup is \{$I$, $X_1Z_2Z_3X_4$, $X_2Z_3Z_4X_5$, $X_1X_3Z_4Z_5$, $Z_1X_2X_4Z_5$\}, and the code words are,
\be
|0_\mathcal{L}\rangle&=&\frac{1}{4}(IIIII+X_1Z_2Z_3X_4+X_2Z_3Z_4X_5\non\\
&&+X_1X_3Z_4Z_5+Z_1X_2X_4Z_5+X_1Y_2Y_4X_5\non\\
&&+Z_2Y_3Y_4Z_5+Y_1Y_2Z_3Z_5+X_1X_2Y_3Y_5\non\\
&&+Z_1Z_3Y_4Y_5+Y_1X_2X_3Y_4+Y_2X_3X_4Y_5\non\\
&&+Y_1Z_2Z_4Y_5+Y_1Y_2X_4X_5+Z_1Y_2Y_3Z_4\non\\
&&+Z_1Z_2X_3X_5)|00000\rangle\non\\
&=&\frac{1}{4}[|00000\rangle+|10010\rangle+|01001\rangle+|10100\rangle\non\\
&&+|01010\rangle-|11011\rangle-|00110\rangle-|11000\rangle\non\\
&&-|11101\rangle-|00011\rangle-|11110\rangle-|01111\rangle\non\\
&&-|10001\rangle-|01100\rangle-|10111\rangle+|00101\rangle],\non\\
|1_\mathcal{L}\rangle&=&\frac{1}{4}(IIIII+X_1Z_2Z_3X_4+X_2Z_3Z_4X_5\non\\
&&+X_1X_3Z_4Z_5+Z_1X_2X_4Z_5+X_1Y_2Y_4X_5\non\\
&&+Z_2Y_3Y_4Z_5+Y_1Y_2Z_3Z_5+X_1X_2Y_3Y_5\non\\
&&+Z_1Z_3Y_4Y_5+Y_1X_2X_3Y_4+Y_2X_3X_4Y_5\non\\
&&+Y_1Z_2Z_4Y_5+Y_1Y_2X_4X_5+Z_1Y_2Y_3Z_4\non\\
&&+Z_1Z_2X_3X_5)|11111\rangle\non\\
&=&\frac{1}{4}[|11111\rangle+|01101\rangle+|10110\rangle+|01011\rangle\non\\
&&+|10101\rangle-|00100\rangle-|11001\rangle-|00111\rangle\non\\
&&-|00010\rangle-|11100\rangle-|00001\rangle-|10000\rangle\non\\
&&-|01110\rangle-|10011\rangle-|01000\rangle+|11010\rangle].\non
\ee

For the seven-qubit Steane code, the subgroup is \{$I$, $X_4X_5X_6X_7$, $X_2X_3X_6X_7$, $X_1X_3X_5X_7$\}, and the code words are,
\be
|0_\mathcal{L}\rangle&=&\frac{1}{\sqrt{8}}(IIIIIII+X_1X_3X_5X_7+X_2X_3X_6X_7\non\\
&&+X_1X_2X_5X_6+X_4X_5X_6X_7+X_1X_3X_4X_6\non\\
&&+X_2X_3X_4X_5+X_1X_2X_4X_7)|0000000\rangle\non\\
&=&\frac{1}{\sqrt{8}}[|0000000\rangle+|1010101\rangle\non\\
&&+|0110011\rangle+|1100110\rangle+|0001111\rangle\non\\
&&+|1011010\rangle+|0111100\rangle+|1101001\rangle],\non\\
|1_\mathcal{L}\rangle&=&\frac{1}{\sqrt{8}}(IIIIIII+X_1X_3X_5X_7+X_2X_3X_6X_7\non\\
&&+X_1X_2X_5X_6+X_4X_5X_6X_7+X_1X_3X_4X_6\non\\
&&+X_2X_3X_4X_5+X_1X_2X_4X_7)|1111111\rangle\non\\
&=&\frac{1}{\sqrt{8}}[|1111111\rangle+|0101010\rangle\non\\
&&+|1001100\rangle+|0011001\rangle+|1110000\rangle\non\\
&&+|0100101\rangle+|1000011\rangle+|0010110\rangle].\non
\ee

For the nine-qubit case, the known Shor code is constructed as shown in Fig.~\ref{figure1}. If one nine-qubit code constructed similarly with the codes above, the subgroup is \{$I$, $X_1X_2X_3X_4X_5X_6$, $X_4X_5X_6X_7X_8X_9$\}, and the code words are,
\be
|0_\mathcal{L}\rangle&=&\frac{1}{2}(IIIIIIIII+X_1X_2X_3X_4X_5X_6\non\\
&&+X_4X_5X_6X_7X_8X_9+X_1X_2X_3X_7X_8X_9)\non\\
&&|000000000\rangle\non\\
&=&\frac{1}{2}[|000000000\rangle+|111111000\rangle\non\\
&&+|000111111\rangle+|111000111\rangle],\non\\
|1_\mathcal{L}\rangle&=&\frac{1}{2}(IIIIIIIII+X_1X_2X_3X_4X_5X_6\non\\
&&+X_4X_5X_6X_7X_8X_9+X_1X_2X_3X_7X_8X_9)\non\\
&&|111111111\rangle\non\\
&=&\frac{1}{2}[|111111111\rangle+|000000111\rangle\non\\
&&+|111000000\rangle+|000111000\rangle].\non
\ee

\end{appendix}

\section{appendix b: The errors can be recovered for the code in Eq.~(\ref{e2})}
\begin{appendix}
\label{a2}

(I) No error occurred, through 3 series inner measurements and the outer measurement, the results $S_i=1,1,i=1,2,...9$, and $R_{f(i)}=1,1,f=Ceil-Integer[\frac{i}{3}]$. Here, $S$ is the result for the inner measurement, and $R$ is the result for the outer measurement.

(II) Only one-qubit error occurred on qubit-$i$ ($i=1,2,...9$), and there are 3 cases. For convenience, we define the measurement results $1,-1/-1,1/-1,-1\equiv ``-"$, and $1,1\equiv ``+"$.
(1) Bit-flip error occurred on qubit-$i$, $S_i=-$, and $R_{f(i)}=+$;

(2) Phase-flip error occurred on qubit-$i$, $S_i=+$, and $R_{f(i)}=-$.

(3) Bit-phase-flip error occurred on qubit-$i$, $S_i=-$, and $R_{f(i)}=-$. Specifically, $R_{f(i)}=-1,1,f(i)=1$, $R_{f(i)}=1,-1,f(i)=3$, and $R_{f(i)}=-1,-1,f(i)=2$, and the 3 cases can be recovered.

(III) Two-qubit error occurred on qubit-$i$ and qubit-$j$, no error occurred on the rest outer qubit-$m$ ($f(i)\neq f(j)\neq f(m),i,j,m=1,2,...9$), and there are 6 cases.

(1) Bit-flip error occurred on qubit-$i$ and qubit-$j$,
\be
S_{i}=-,S_{j}=-,S_{m}=+,R=+; \non
\ee

(2) Bit-flip error occurred on qubit-$i$, and phase-flip error occurred on qubit-$j$,
\be
S_{i}=-,S_{j}=S_{m}=+,R=-; \non
\ee

(3) Bit-flip error occurred on qubit-$i$, and bit-phase-flip error occurred on qubit-$j$,
\be
S_{i}&=&-,S_{j}=-,S_{m}=+,R=-; \non
\ee

(4) Phase-flip error occurred on qubit-$i$ and qubit-$j$,
\be
S_{i}=S_{j}=S_{m}=+,R=-; \non
\ee
It conflicts with the case phase-flip error occurred on qubit-$m$, which corresponding to the one-qubit case `(2)'.

(5) Phase-flip error occurred on qubit-$i$, and bit-phase-flip error occurred on qubit-$j$,
\be
S_{j}&=&-,S_{i}=S_{m}=+,R=-; \non
\ee
It conflicts with the case bit-flip error occurred on qubit-$j$, and phase-flip error occurred on qubit-$m$, which corresponding to the two-qubit case `(2)'.

(6) Bit-phase-flip error occurred on qubit-$i$ and qubit-$j$,
\be
S_{i}&=&-,S_{j}=-,S_{m}=+,R=-. \non
\ee

(IV) Three-qubit error occurred on qubit-$i$, qubit-$j$ and qubit-$m$ ($f(i)\neq f(j)\neq f(m),i,j,m=1,2,...9$), and there are 9 cases.

(1) Bit-flip error occurred on qubit-$i$, qubit-$j$ and qubit-$m$,
\be
S_{i}&=&-,S_{j}=-,S_{m}=-,R=+; \non
\ee

(2) Bit-flip error occurred on qubit-$i$ and qubit-$j$, and phase-flip error occurred on qubit-$m$,
\be
S_{i}&=&-,S_{j}=-,S_{m}=+,R=-; \non
\ee
It conflicts with the case bit-phase-flip error occurred on qubit-$i$ and qubit-$j$, which corresponding to the two-qubit case `(6)'.

(3) Bit-flip error occurred on qubit-$i$ and qubit-$j$, and bit-phase-flip error occurred on qubit-$m$,
\be
S_{i}&=&-,S_{j}=-,S_{m}=-,R=-; \non
\ee

(4) Phase-flip error occurred on qubit-$i$, qubit-$j$, and qubit-$m$,
\be
S_{i}&=&S_{j}=S_{m}=+,R=+; \non
\ee
It conflicts with the case no error occurred.

(5) Phase-flip error occurred on qubit-$i$ and qubit-$j$, and bit-flip error occurred on qubit-$m$,
\be
S_{i}&=&S_{j}=+,S_{m}=-,R=-; \non
\ee
It conflicts with the case bit-phase-flip error occurred on qubit-$m$, which corresponding to the one-qubit case `(3)'.

(6) Phase-flip error occurred on qubit-$i$ and qubit-$j$, and bit-phase-flip error occurred on qubit-$m$,
\be
S_{i}&=&S_{j}=+,S_{m}=-,R=+; \non
\ee
It conflicts with the case bit-flip error occurred on qubit-$m$, which corresponding to the one-qubit case `(1)'.

(7) Bit-phase-flip error occurred on qubit-$i$, qubit-$j$, and qubit-$m$,
\be
S_{i}&=&-,S_{j}=-,S_{m}=-,R=+; \non
\ee
It conflicts with the case bit-flip error occurred on qubit-$i$, qubit-$j$, and qubit-$m$, which corresponding to the three-qubit case `(1)'.

(8) Bit-phase-flip error occurred on qubit-$i$ and qubit-$j$, and bit-flip error occurred on qubit-$m$,
\be
S_{i}&=&-,S_{j}=-,S_{m}=-,R=-; \non
\ee
It conflicts with the case bit-flip error occurred on qubit-$i$ and qubit-$j$, and bit-phase-flip error occurred on qubit-$m$, which corresponding to the three-qubit case `(3)'.

(9) Bit-phase-flip error occurred on qubit-$i$ and qubit-$j$, and phase-flip error occurred on qubit-$m$,
\be
S_{i}&=&-,S_{j}=-,S_{m}=+,R=+. \non
\ee
It conflicts with the case bit-flip error occurred on qubit-$i$ and qubit-$j$, which corresponding to the two-qubit case `(1)'.

We can only keep one of the conflicting cases for recovering, and the selection criteria is depending on the weight of errors. And this code is more suitable for the high-wight bit-flip noise.
\end{appendix}

\section{appendix c: The errors can be recovered for the code in Eq.~(\ref{e4})}
\begin{appendix}
\label{a3}

(I) No error occurred, through 3 series inner measurements and the outer measurement, the results $S_i=1,1,i=1,2,...9$, and $R_{f(i)}=1,1,f=Ceil-Integer[\frac{i}{3}]$. Here, $S$ is the result for the inner measurement, and $R$ is the result for the outer measurement.

(II) Only one-qubit error occurred on qubit-$i$ ($i=1,2,...9$), and there are 3 cases.

(1) Phase-flip error occurred on qubit-$i$, $S_i=-$, and $R_{f(i)}=+$;

(2) Bit-flip error occurred on qubit-$i$, $S_i=+$, and $R_{f(i)}=-$.

(3) Bit-phase-flip error occurred on qubit-$i$, $S_i=-$, and $R_{f(i)}=-$. Specifically, $R_{f(i)}=-1,1,f(i)=1$, $R_{f(i)}=1,-1,f(i)=3$, and $R_{f(i)}=-1,-1,f(i)=2$, and the 3 cases can be recovered.

(III) Two-qubit error occurred on qubit-$i$ and qubit-$j$, no error occurred on the rest outer qubit-$m$ ($f(i)\neq f(j)\neq f(m),i,j,m=1,2,...9$), and there are 6 cases.

(1) Phase-flip error occurred on qubit-$i$ and qubit-$j$,
\be
S_{i}=-,S_{j}=-,S_{m}=+,R=+; \non
\ee

(2) Phase-flip error occurred on qubit-$i$, and bit-flip error occurred on qubit-$j$,
\be
S_{i}=-,S_{j}=S_{m}=+,R=-; \non
\ee

(3) Phase-flip error occurred on qubit-$i$, and bit-phase-flip error occurred on qubit-$j$,
\be
S_{i}&=&-,S_{j}=-,S_{m}=+,R=-; \non
\ee

(4) Bit-flip error occurred on qubit-$i$ and qubit-$j$,
\be
S_{i}=S_{j}=S_{m}=+,R=-; \non
\ee
It conflicts with the case bit-flip error occurred on qubit-$m$, which corresponding to the one-qubit case `(2)'.

(5) Bit-flip error occurred on qubit-$i$, and bit-phase-flip error occurred on qubit-$j$,
\be
S_{j}&=&-,S_{i}=S_{m}=+,R=-; \non
\ee
It conflicts with the case phase-flip error occurred on qubit-$j$, and bit-flip error occurred on qubit-$m$, which corresponding to the two-qubit case `(2)'.

(6) Bit-phase-flip error occurred on qubit-$i$ and qubit-$j$,
\be
S_{i}&=&-,S_{j}=-,S_{m}=+,R=-. \non
\ee

(IV) Three-qubit error occurred on qubit-$i$, qubit-$j$ and qubit-$m$ ($f(i)\neq f(j)\neq f(m),i,j,m=1,2,...9$), and there are 9 cases.

(1) Phase-flip error occurred on qubit-$i$, qubit-$j$ and qubit-$m$,
\be
S_{i}&=&-,S_{j}=-,S_{m}=-,R=+; \non
\ee

(2) Phase-flip error occurred on qubit-$i$ and qubit-$j$, and bit-flip error occurred on qubit-$m$,
\be
S_{i}&=&-,S_{j}=-,S_{m}=+,R=-; \non
\ee
It conflicts with the case bit-phase-flip error occurred on qubit-$i$ and qubit-$j$, which corresponding to the two-qubit case `(6)'.

(3) Phase-flip error occurred on qubit-$i$ and qubit-$j$, and bit-phase-flip error occurred on qubit-$m$,
\be
S_{i}&=&-,S_{j}=-,S_{m}=-,R=-; \non
\ee

(4) Bit-flip error occurred on qubit-$i$, qubit-$j$, and qubit-$m$,
\be
S_{i}&=&S_{j}=S_{m}=+,R=+; \non
\ee
It conflicts with the case no error occurred.

(5) Bit-flip error occurred on qubit-$i$ and qubit-$j$, and phase-flip error occurred on qubit-$m$,
\be
S_{i}&=&S_{j}=+,S_{m}=-,R=-; \non
\ee
It conflicts with the case bit-phase-flip error occurred on qubit-$m$, which corresponding to the one-qubit case `(3)'.

(6) Bit-flip error occurred on qubit-$i$ and qubit-$j$, and bit-phase-flip error occurred on qubit-$m$,
\be
S_{i}&=&S_{j}=+,S_{m}=-,R=+; \non
\ee
It conflicts with the case phase-flip error occurred on qubit-$m$, which corresponding to the one-qubit case `(1)'.

(7) Bit-phase-flip error occurred on qubit-$i$, qubit-$j$, and qubit-$m$,
\be
S_{i}&=&-,S_{j}=-,S_{m}=-,R=+; \non
\ee
It conflicts with the case phase-flip error occurred on qubit-$i$, qubit-$j$, and qubit-$m$, which corresponding to the three-qubit case `(1)'.

(8) Bit-phase-flip error occurred on qubit-$i$ and qubit-$j$, and phase-flip error occurred on qubit-$m$,
\be
S_{i}&=&-,S_{j}=-,S_{m}=-,R=-; \non
\ee
It conflicts with the case phase-flip error occurred on qubit-$i$ and qubit-$j$, and bit-phase-flip error occurred on qubit-$m$, which corresponding to the three-qubit case `(3)'.

(9) Bit-phase-flip error occurred on qubit-$i$ and qubit-$j$, and bit-flip error occurred on qubit-$m$,
\be
S_{i}&=&-,S_{j}=-,S_{m}=+,R=+. \non
\ee
It conflicts with the case phase-flip error occurred on qubit-$i$ and qubit-$j$, which corresponding to the two-qubit case `(1)'.

We can only keep one of the conflicting cases for recovering, and the selection criteria is depending on the weight of errors. And this code is more suitable for the high-wight phase-flip noise.
\end{appendix}

\end{document}